\documentclass[12pt,aps,]{article}
\usepackage{amsmath}
\usepackage{amssymb}
\usepackage{amsfonts}
\usepackage{hyphenat}
\usepackage{verbatim}

\begin{document}

\title{On the momentum of solitons and vortex rings in a superfluid}
\author{L.P. Pitaevskii \\
P.L. Kapitza Institute for Physical Problems RAS,\\
119334, Moscow, ul. Kosygina 2, RF;\\
Dipartimento di Fisica, Universit{\`{a}} di Trento and INO--CNR BEC Center,\\
I--38123 Trento, Povo, Italy. \\
}
\date{}
\maketitle

\begin{abstract}
This paper is devoted to the calculation of the momentum of localized
excitations, such as solitons and vortex rings, moving in a superfluid. The
direct calculation of the momentum by integration of the mass flux density
results in a badly--converging integral. I suggest a method for the
renormalization of the integral with the explicit separation of a term
related to the vortex line. This term can be calculated explicitly and gives
the main contribution for the rings whose size is large compared to the
healing length. I compare my method with the Jones and Roberts prescription
for renormalization. I investigate the case of a uniform superfluid, and
that of a superfluid in a cylindrical trap. I discuss the calculation of the
jump in the phase of the order parameter and obtain a simple estimate for
this jump for a large ring in the trap.
\end{abstract}


\section{Introduction}

One of the characteristic properties of a superfluid is the possible
existence of localized stationary excitations. These can be solitons, vortex
rings, solitonic vortices, or other more complicated objects. The study of
the vortex rings has resulted in important progress in the understanding of
the physics of superfluid helium. A very interesting possibility arose after
the creation of new superfluid systems like ultracold Bose and Fermi gases,
which are confined in traps. The dynamics of the excitations in these
systems is an important theoretical and experimental problem due to the
presence of external fields. At the same time, due to the diluteness of the
gases, one can develop a microscopic theory which is sufficiently detailed.
If the trapping field changes slowly enough, it is convenient to solve the
problem in two steps. As a first step, it is reasonable to find the energy
and momentum for an excitation in a uniform fluid. (In the case of an
elongated trap, the fluid is uniform in the axial direction.) Second, one
can analyze the motion of the excitations using the semi--classical
equations of motion. Depending on the nature of the problem, the energy
should be expressed in terms of momentum or velocity. However, a difficulty
arises which had already been encountered in classical hydrodynamics (see
for instance \cite[\textsection 11]{LL6} or \cite[ch.~6]{Bat73}). 
If one tries to calculate the momentum by integration of the mass flux
density $\mathbf{j}$ over the volume of a fluid, i.e. using the equation: 
\begin{equation}
\mathbf{p}=\int \mathbf{j} \, d^{3}x\;,  \label{pj}
\end{equation}%
then, for stationary motion, the integrand does not decrease fast enough and
the value of the integral depends on the shape of the integration domain.
Different prescriptions have been suggested to overcome this difficulty. If
one knows the energy $\varepsilon$ as a function of the excitation velocity $%
V$, the momentum can be calculated by integrating the following equation: 
\begin{equation}
\left( \partial \varepsilon /\partial p\right) =V\;.  \label{ham}
\end{equation}%
This method was used in the Landau--Lifshitz book \cite{LL6}. However, it
does not allow for the definition of the integration constant, which is
important for the calculation of the Landau critical velocity. In his book 
\cite{Bat73}, Batchelor assumed that the moving body was initially at rest
and subsequently set in motion by an external force, and he calculated the
momentum which was transferred to the body and the fluid (see ch.~7). The
same method was used to calculate the momentum of a classical vortex ring.
Roberts and Grant \cite{RG71}, and Jones and Roberts \cite{JR82}, used a
similar approach to calculate the momentum of a vortex ring in a
Bose--Einstein condensate (BEC), using the Gross--Pitaevskii (GP) equation.
In Ref.~\cite{RG71}, the authors considered a ring whose radius is large
compared to the healing length $\xi$. In Ref.~\cite{JR82}, a ring with
arbitrary radius was considered, and a general renormalized equation for the
momentum was obtained in the form of an absolutely convergent integral. I
shall discuss it below. The existence of such an expression is related to
the fact that a perturbation propagates in a fluid with a finite velocity,
which does not exceed the velocity of sound. Because the excitation was
formed at some initial time, this means that at large enough distances the
velocity of flow is zero. There is, however, a difficulty, because at these
distances the problem is not stationary. In this article I suggest a
different renormalized equation for the momentum. This equation is
convenient for the calculation of the momentum of an excitation which
contains a quantized vortex of large size, because the main term related to
the vorticity is separated explicitly. I shall assume that the superfluid is
described by an order parameter $\Psi$. Its phase $\phi$ defines the
superfluid velocity according to the equation $\mathbf{v}=\frac{\hbar}{M}%
\mathbf{\nabla }\phi$, where $M=m$ for a bosonic superfluid and $M=2m$ for a
fermionic superfuid, and $m$ is the atomic mass. For simplicity, I shall
also assume that the mass flux density is equal to 
\begin{equation}
\mathbf{j}=\rho \mathbf{v}\;,  \label{j}
\end{equation}%
with $\rho$ being the mass density of the fluid. This equation is licit in
the context of the GP equation and several of it generalizations. (In the
general case, $\mathbf{j}$ can also depend on the derivatives of $\mathbf{v}$%
.) However, this last assumption is not necessary. It is only important that
Eq.~(\ref{j}) be always satisfied at large distances, where hydrodynamics is
valid. One can present the density as $\rho =\rho _{\infty }+\delta \rho ,$
where $\rho _{\infty }$ is the unperturbed density at large distances. Then
the momentum will be presented as: 
\begin{equation}
\mathbf{p}=\frac{\hbar }{M}\int \rho _{\infty }\mathbf{\nabla }\phi d^{3}x+%
\frac{\hbar }{M}\int \delta \rho \mathbf{\nabla }\phi d^{3}x\;.  \label{p}
\end{equation}%
The second integral converges absolutely, because the density perturbation $%
\delta \rho$ , being quadratic in $\mathbf{v}$, decreases fast. The first
term can be calculated in a general form, taking into account that the phase
is equal to zero at infinity. Note that this method was used in \cite[%
\textsection 65]{LL6} for the calculation of the momentum of a wave packet
of sound waves. I shall consider below several examples of the application
of Eq.~(\ref{p}) in order to demonstrate the effectiveness of the method.

\section{An instructive example: Induced momentum of a sphere moving in a
fluid.}

As a first example, I shall consider the classical problem of the
calculation of the momentum of a sphere of radius $R$ moving with the
velocity $\mathbf{V}$ in an incompressible liquid. We shall see that the
problem can be solved by the suggested method in a surprisingly simple way.
The second term in Eq.~(\ref{p}) is equal to zero due to the
incompressibility of the liquid. The first term can be transformed into an
integral on the infinitely remote surface, which is equal to zero, and an
integral over the surface of the sphere. As a result one gets for the
momentum of the liquid: 
\begin{equation}
\mathbf{p}_{fl}=-\rho \frac{\hbar }{M}\oint \mathbf{n}\phi d\sigma \;,
\label{main1}
\end{equation}%
where the vector $\mathbf{n}$ is normal to the surface and $d\sigma$ is an
element of the surface. (It is called the induced momentum.) Note that the
integral over the remote surface being zero implies that the liquid is still
compressible and that the remote surface is on the distances $r>ct$, where $c
$ is the speed of sound, $t$ is the time from the initial moment of motion.
However, if the velocity of motion is much smaller than the speed of sound, $%
V\ll c$ , then one can neglect the compressibility on distances of the order
of $R$ and a solution of the hydrodynamic problem gives the potential for
the velocity on the surface: 
\begin{equation}
\frac{\hbar }{M}\phi =-\frac{R}{2}\mathbf{n\cdot V}\;.
\end{equation}%
In this case, the momentum is 
\begin{equation}
\mathbf{p}_{fl}=\frac{2\pi }{3}R^{3}\rho \mathbf{V} \; ,
\end{equation}%
in accordance with Eq.~(11.6) in \cite{LL6} (see also Eq.~(6.4.29) in \cite%
{Bat73}).

\section{Vortex rings in a uniform fluid}

Let us now consider an excitation moving with a constant velocity $\mathbf{V}
$ in an infinite volume of a uniform fluid. Consider first an excitation
which does not contain vortex lines. For instance, it can be the wonderful
Jones--Roberts soliton. A vortex ring degenerates into this excitation when
its velocity approaches $V=0.88c$ \cite{JR82}. (In the 2D case an analytical
solitonic solution was constructed by Manakov et al \cite{M77}.) Then the
first integral in Eq.~(\ref{p}) can be transformed in a surface integral
over the remote surface, which is equal to zero, and the momentum of the
excitation is: 
\begin{equation}
\mathbf{p}=\frac{\hbar }{M}\int \delta \rho \mathbf{\nabla }\phi d^{3}x\;.
\label{p2}
\end{equation}%
However, if there is a loop of a quantum vortex in the structure of the
excitation, the situation is different. In this case, the phase $\phi$ is
not a single--valued function of the coordinates: it acquires an additional $%
2\pi $ when going around the vortex line. Now, in order to apply the Gauss
theorem, one should perform a cut. The values of $\phi$ on the two sides of
the cut differ by $2\pi$. This makes $\phi$ single--valued. One can choose
as a cut any surface which closes the aperture of the loop. Now, besides the
remote surface, whose contribution is zero, one should integrate over both
sides of the cut. We finally get for the momentum: 
\begin{equation}
\mathbf{p}=\rho _{\infty }\frac{2\pi \hbar }{M}\oint \mathbf{n}d\sigma +%
\frac{\hbar }{M}\int \delta \rho \mathbf{\nabla }\phi d^{3}x\;.  \label{main}
\end{equation}%
The integration in the first term is performed over the surface of the cut,
the vector $\mathbf{n}$ is normal to this surface, and $d\sigma$ is an
element of the surface. This equation is the main result of the present
paper. In the case of a circular ring, moving along the $x$ direction, the
momentum is:%
\begin{equation}
p=\rho _{\infty }\frac{2\pi ^{2}\hbar }{M}R^{2}+\frac{\hbar }{M}\int \delta
\rho \mathbf{\nabla }_{x}\phi d^{3}x\;.  \label{p3}
\end{equation}%
An important property of this expression is that for a large ring, whose
radius $R$ satisfies 
\begin{equation}
R\gg \xi \; ,  \label{Rxi}
\end{equation}
with $\xi \sim \hbar /mc$ being the healing length of the fluid, the two
terms in Eq.~(\ref{p3}) have different orders of magnitude. Indeed, a simple
estimate shows that the second term is of the order of $\rho _{\infty}\hbar
\xi ^{2}/M$, that is, much smaller than the first term\footnote{%
It is important for the estimate that, under the condition (\ref{Rxi}), the
velocity of the ring $V \ll c$.}. Thus, the momentum of the large ring is
given by the first term in Eq.~(\ref{p3}). This equation is well known. It
was used to describe the hydrodynamics of a classical liquid. Feynman used
it when discussing the critical velocity related to the creation of the
vortex rings \cite{F55}. However, this equation is difficult to derive
starting from a microscopic theory such as the GP equation (see the
discussion in \cite{RG71}). Using the suggested method, this problem is
easily solved. Jones and Roberts used a different method to renormalize the
equation for the momentum of a vortex ring of arbitrary radius on the basis
of the GP equation. They subtracted the integral 
\begin{equation}
\frac{\hbar }{2iM}\sqrt{\rho _{\infty }}\int \left[ \mathbf{\nabla }\Psi -%
\mathbf{\nabla }\Psi ^{\ast }\right] d^{3}x=-\frac{\hbar }{M}\sqrt{\rho
_{\infty }}\int \mathbf{\nabla }\left( \sqrt{\rho }\sin \phi \right) d^{3}x
\label{JR}
\end{equation}%
from the general equation (\ref{pj}) for the momentum. (Recall that, in the
GP equation, $\Psi =\sqrt{\rho }e^{i\phi }$.) This integral is equal to
zero, because, unlike $\phi $, the order parameter $\Psi $ is a
single--valued function of the coordinates, and the integral is transformed
into an integral over the remote surface only, where $\phi=0$. They finally
obtained the following expression for the momentum: 
\begin{equation}
\mathbf{p}=\frac{\hbar }{2iM}\int \left[ \left( \Psi ^{\ast }-\sqrt{\rho
_{\infty }}\right) \mathbf{\nabla }\Psi -\left( \Psi -\sqrt{\rho _{\infty }}%
\right) \mathbf{\nabla }\Psi ^{\ast }\right] d^{3}x\;.  \label{pJR}
\end{equation}%
The authors of Ref.~\cite{KP03} used this equation to calculate the momenta
of solitons, vortex rings and solitonic vortices. The integral (\ref{pJR})
is absolutely convergent, but it is difficult to simplify this equation in
the case of a large ring, because one should find a corresponding asymptotic
equation for $\Psi$. An approximate wave function which is accurate enough
was constructed by Berloff, see \cite[Eq.~25]{B04}. Using this function, one
can calculate the momentum analytically according to Eq.~(\ref{pJR}). In the
limit $R\rightarrow \infty$, one gets $\rho _{\infty }2\pi ^{2}\hbar R^{2}/M$
\cite{Berl}, in accordance with Eq.~(\ref{p3}), as it should be.

\section{Solitons and vortex rings in a cylindrical trap.}

A natural method for the experimental investigation of the dynamics of the
excitations is the observation of their motion along the axis of an
elongated trap. As I have explained above, the first step to develop the
theory of such a motion is to calculate the energy and the momentum of the
excitation in the trap. My goal is to calculate the momentum. I shall use
the same considerations as in the case of a uniform fluid. However, the
situation in a cylindrical trap has its peculiarities. For simplicity, I
shall consider a fluid in an axially--symmetric potential. Let us call $x$
the coordinate along the axis and $r$ the radial coordinate. It is not
difficult to show that the velocity field decreases fast enough at large
distances $\left\vert x\right\vert $ from any stationary excitation. Thus
the integral in Eq.~(\ref{pj}) converges absolutely. Let us call $p_{s}$ its
result. It seems that there no longer is any motivation for the
renormalization of the momentum. However, this is not the case. The point is
that the stationary solution for the order parameter has a phase jump
between $x=-\infty $ and $x=+\infty $: 
\begin{equation}
\Delta \phi =\phi \left( x=\infty \right) -\phi \left( x=-\infty \right) \;.
\label{deltaphi}
\end{equation}%
Nevertheless, for an excitation created at $t=0$, there is no perturbation
of the phase at infinity. This means that, at large enough distances from
the excitation, there arises a counter--current which compensates the phase
jump. If the region of the counterflow is long enough, the velocity $\bar{v}$
of the fluid in the counterflow is small and the counterflow does not
contribute to the energy, but it does contributes to the momentum\footnote{%
It is possible to show (see \cite{KP03}) that $\phi (\pm \infty )$ does not
depend on $r$. Correspondingly, the velocity $\bar{v}$ does not depend on $r$
either.}. Let $\rho _{\infty }(r)$ denote the density of the fluid at large
distances $\left\vert x\right\vert \rightarrow \infty $. Then, the
contribution of the counterflow to the momentum is: 
\begin{equation}
\Delta p=\int \rho _{\infty }2\pi rdr\frac{\hbar }{M}\int \partial _{x}\phi
dx=-\hbar \rho _{1\infty }\Delta \phi /M\;,  \label{Dp}
\end{equation}%
where the integral with respect to $x$ is taken over the region of the
counterflow, and $\rho _{1\infty }=\int \rho _{\infty }2\pi rdr$ is the 1D
density of the unperturbed fluid. Note that the result (\ref{Dp}) does not
depend on the velocity distribution in the counterflow region or on its
length. Thus the momentum is \cite{SD11}: 
\begin{equation}
p=p_{s}+\Delta p=\frac{\hbar }{M}\int \rho \left( x,r\right) \partial
_{x}\phi d^{3}x-\rho _{1\infty }\frac{\hbar }{M}\Delta \phi \;.  \label{p4}
\end{equation}%
In this equation, the quantities $\rho $ and $\phi $ should be calculated
assuming that they depend on $x$ and $t$ through the combination $x-Vt$
(stationary motion). The integral over $x$ is taken from $-\infty $ to $%
\infty $. As I have already said, it converges under the accepted
assumption. Note that adding to the phase an integer multiple of $2\pi $
does not change the state of the system. This means that values of the
momentum which differ by a multiple of $2\pi \hbar \rho _{1\infty }/M$ are
physically equivalent. This is a general property of a quantum fluid in a 1D
geometry. This fluid can be considered as a 1D periodic structure with the
period $M/\rho _{1\infty }$, smeared by quantum fluctuations, and $p$ has
properties of the quasimomentum \cite{H81}. One can obtain a different
expression for the momentum, using the general Eq.~(\ref{p}). Calculating
the surface integral as in the previous section, we find for a circular
vortex ring: 
\begin{equation}
p=\frac{2\pi \hbar }{M}\int_{0}^{R}\rho _{\infty }(r)2\pi rdr+\frac{\hbar }{M%
}\int \delta \rho \mathbf{\nabla }_{x}\phi d^{3}x\;.  \label{p4a}
\end{equation}%
An advantage of this equation, analogously to equation (\ref{p3}), is that
in the case of a large ring, $R\gg \xi $, the first term of this equation is
much larger than the second, which can be neglected. (This situation occurs
only in the Thomas--Fermi limit, where $R_{\perp }\gg \xi $, with $R_{\perp }
$ being the radial size of the cloud.) Equation (\ref{p4a}) was derived and
used in my paper \cite{LP13}.

\section{Estimate of the phase jump.}

The phase jump $\Delta \phi$ which we considered in the previous section is
an interesting physical quantity. It can be measured by observing the
interference pattern in the fluid expansion which occurs after the trap is
switched off. Of course, if one can calculate the order parameter around the
excitation, one will know the phase jump. However, it is interesting to
estimate the jump for a large--sized vortex ring, that is, $R\gg \xi $, in a
cylindrical trap. Let us equalize the equations (\ref{p4}) and (\ref{p4a})
for the momentum: 
\begin{equation}
\Delta \phi =-\frac{2\pi }{\rho _{1}}\int_{0}^{R}\rho _{\infty }(r)2\pi rdr-%
\frac{1}{\rho _{1}}\int \delta \rho \mathbf{\nabla }_{x}\phi d^{3}x+\frac{M}{%
\hbar }p_{s}\;.  \label{deltaphi1}
\end{equation}%
One can transform the quantity $p_{s}$ in this equation exploiting Galilean
invariance. Let us consider the mass flux density integrated over the
cross--section of the trap, $j_{1}^{\left( 0\right) }=\int j_{x}dydz$, in
the coordinate frame moving with the velocity of the ring. In this frame,
the density of the fluid does not depend on time. Then, from the continuity
equation $\partial _{t}\rho _{1}+\partial _{x}j_{1}^{\left( 0\right) }=0$,
it follows that $j_{1}^{\left( 0\right) }$ does not depend on the coordinate
and is equal to its value at large distances, where the 1D density is equal
to the unperturbed value $\rho _{1\infty }$ and the velocity of fluid is $-V$%
. Thus $j_{1}^{\left( 0\right) }=-$ $\rho _{1\infty }V$. Now, the Galilean
transformation gives $j_{1}(x)=j_{1}^{\left( 0\right) }+\rho _{1}\left(
x\right) V=\delta \rho _{1}V$ for the current density in the laboratory
frame. Integration with respect to $x$ yields $p_{s}=V\int \delta \rho_{1}dx$%
. The quantity $M_{s}=\int \delta \rho _{1}dx$ represents the mass of the
excitation. Substituting this expression into Eq.~(\ref{deltaphi1}) gives $%
\Delta \phi$: 
\begin{equation}
\Delta \phi =-\frac{2\pi }{\rho _{1}}\int_{0}^{R}\rho _{\infty }(r)2\pi rdr+%
\frac{1}{\rho _{1}}\int \delta \rho \left( \frac{M}{\hbar }V-\mathbf{\nabla }%
_{x}\phi \right) d^{3}x\;.  \label{deltaphi2}
\end{equation}%
It is obvious that for a large ring the second term in this equation is
small and can be neglected. Then, the phase jump is given by the simple
equation: 
\begin{equation}
\Delta \phi \approx -2\pi \frac{\rho _{1R}}{\rho _{1}}\;,  \label{DF}
\end{equation}%
where $\rho _{1R}=\int_{0}^{R}\rho _{\infty }(r)2\pi rdr$ is the 1D fluid
density contained in a cylinder of radius $R$. It would be interesting to
check this equation in numerical simulations of the GP equation, analogous
to \cite{KP03}, but for a vortex ring with $R_{\perp }\gg \xi $. Note that
it is not difficult to generalize the derived equations for a geometry which
does not present axial symmetry, when in the trap there exists a vortex line
which ends on the border of the sample. (I am considering the Thomas--Fermi
limit, in which case the boundary is sharp.) In this case the cut should be
drawn from the line to the boundary. Such a situation occurs for the
solitonic vortex \cite{KP03,BR02,Z14,D14}.

In conclusion, I have presented in this paper a convenient method for the
calculation of the momentum of localized excitations in superfluids. The
method reduces the problem to an absolutely--convergent integral, and it is
particularly convenient in the presence of vortex lines of large length.

I thank P.H. Roberts for useful comments on the paper \cite{JR82},
N.G.~Berloff for discussions and the calculation of momentum according to
the wave function \cite{B04}, S.~Komineas for sending data from the
calculations of Ref.~\cite{KP03}, D.J.~Papoular, T.~Yefsah and I.A. Fomin
for discussions.

This work was supported by the ERC through the QGBE grant, the Provincia
Autonoma di Trento, and the Italian MIUR through the PRIN--2009 grant.

I am glad that the present paper will be published in the special issue of
JETP devoted to the 75--th birthday of A.F.~Andreev. I have had the
possibility to familiarize myself with the work of this outstanding
physicist from the very beginning of his scientific carrier. In my
recollection, A.F.~Andreev has succeeded to solve actually any problem he
has tackled. His papers are always based on the deep understanding of the
physical meaning of a phenomenon and extraordinary clearness of thinking.
They are very elegant and have often stimulated new experiments. I always
read his articles with pleasure and great benefit for myself. We are
grateful to A.F.~that he sacrifices a part of his precious time for
scientific management activities, trying as much as he can to save our
science. I congratulate my friend and wish him permanent desire to work and
good health --- to him and his wonderful family. The rest is a matter of
course.

\end{document}